\documentclass[runningheads]{llncs}
\usepackage{graphicx}
\usepackage[T1]{fontenc}
\usepackage{verbatimbox,lipsum}
\usepackage{subcaption}
\usepackage[utf8]{inputenc}
\usepackage{listings}
\usepackage{pifont}
\usepackage{float}
\usepackage{tablefootnote}
\newfloat{lstfloat}{htbp}{lop}
\floatname{lstfloat}{Listing}

\usepackage[most]{tcolorbox}
\usepackage{array}
\usepackage{amsmath}
\begin{document}
%


\title{Managing Comprehensive Research Instrument Descriptions within a Scholarly Knowledge Graph}

\titlerunning{Managing Research Instrument Descriptions within a Knowledge Graph}
%
\author{Muhammad Haris\inst{1}\orcidID{0000-0002-5071-1658} \and
Sören Auer\inst{2,1}\orcidID{0000-0002-0698-2864} \and
Markus Stocker\inst{2,1}\orcidID{0000-0001-5492-3212}
}

\authorrunning{Haris et al.}

\institute{L3S Research Center, Leibniz University Hannover 30167, Hannover, Germany\\
\email {haris@l3s.de} \and
TIB---Leibniz Information Centre for Science and Technology, Germany
\email{\{markus.stocker,auer\}@tib.eu}}

\maketitle              

\begin{abstract}
In research, measuring instruments play a crucial role in producing the data that underpin scientific discoveries. Information about instruments is essential in data interpretation and, thus, knowledge production. However, if at all available and accessible, such information is scattered across numerous data sources. Relating the relevant details, e.g. instrument specifications or calibrations, with associated research assets (data, but also operating infrastructures) is challenging. Moreover, understanding the (possible) use of instruments is essential for researchers in experiment design and execution. To address these challenges, we propose a Knowledge Graph (KG) based approach for representing, publishing, and using information, extracted from various data sources, about instruments and associated scholarly artefacts. The resulting KG serves as a foundation for exploring and gaining a deeper understanding of the use and role of instruments in research, discovering relations between instruments and associated artefacts (articles and datasets), and opens the possibility to quantify the impact of instruments in research.
\end{abstract}
\section{Introduction}
\label{s:introduction}
Continuous advancements in technology have led to the development of specialized instruments, that allow researchers to collect precise and reliable data, leading to significant discoveries in various disciplines including oceanography, seismology, climatology, and botany. For instance, in oceanography, CTD (Conductivity, Temperature, and Depth sensors)~\cite{ctd} measures the salinity, temperature, and depth of seawater; In botany, Chlorophyll afluorescence~\cite{chlorophyll_fluorescence} is useful to assess photosynthetic performance in plants; Seismographs~\cite{balch1984vertical} in seismology provide crucial information on tectonic movements. Despite their significance, the distribution of information about these instruments across disparate sources can make understanding their role in research challenging. A further issue is the inability to unambiguously refer to an instrument, e.g. in metadata describing data or in the literature. Such growth is heterogeneous and unstructured, requires specialized methods to mine and integrate this multidimensional data into structured and machine-understandable form, known as knowledge graph (KG). The information provided by KGs is machine-actionable and could be consumed by different applications to search instruments information in a better way. The integration of data about instruments (e.g., articles and datasets) into a unified knowledge base will facilitate the quick discovery and efficient retrieval.

Knowledge Graph (KG) technology offers a powerful framework for representing information--- in particular also scientific information and knowledge published in the literature--- in a structured manner, enabling efficient search and exploration of data. Several KGs--- or databases, more generally--- have been developed to represent and provide access to (scientific) information in structured form, among others: DBpedia\footnote{\url{https://www.dbpedia.org}}~\cite{dbpedia}, Wikidata\footnote{\url{https://www.wikidata.org/wiki/Wikidata:Main\_Page}}~\cite{wikidata}, Cultural Heritage~\cite{domainspecific}, KnowLife in Life Sciences~\cite{knowlife}, Hi-Knowledge in Invasion Biology\footnote{\url{https://hi-knowledge.org}}~\cite{heger2013conceptual,enders2020conceptual}.

We propose to enrich the existing infrastructure, specifically, the Open Research Knowledge Graph\footnote{\url{https://www.orkg.org/}}~\cite{stocker2023fair}--- with structured scientific knowledge about instruments and their associated artefacts automatically extracted from different data sources. The main purpose of the KG is to capture information about instruments and the digital artefacts produced by these instruments, described in research articles. Most importantly, information about the research work that utilize these instruments to produce the research artefacts which we aim to extract by employing machine learning-based approaches. We summarize the contributions as follows:

\begin{enumerate}
    \item \textit{Mining instruments metadata} from DataCite\footnote{\url{https://datacite.org/}} and Alfred Wegener Institut (AWI)\footnote{\url{https://sensor.awi.de/}} to extract the key information, i.e, instrument name, description, URL and manufacturer.
    
    \item \textit{Retrieve instrument-related artefacts} by analyzing the links that exist between instruments and related artefacts (datasets and articles). We analyze metadata-based links to retrieve datasets produced by instruments as well as articles linked to these datasets. This step allows us to trace research work that mentions the use of instruments and helps determine the context in which these instruments are used.
    
    \item \textit{Discover instruments usage} by analyzing the content of retrieved datasets to determine the parameters measured by the instruments. Additionally, we apply Named Entity Recognition (NER) method to extract additional details from the linked scholarly articles that mention the use of instruments. Specifically, we employ BERT-based models to recognize entities from the articles text, aiming to determine the methods and processes employed in measuring these parameters.
    
    \item \textit{Populate a knowledge graph:} The extracted knowledge about the instruments is represented in a structured format. The knowledge graph represents the relationships between different instruments as well as their properties and applications. The resulting KG can be queried using a SPARQL endpoint to discover relationships among instruments and related digital artefacts.
\end{enumerate}

We address the following research question:
\begin{enumerate}
    \item How can we ensure the discovery of data produced by measuring instruments across various disciplines?
\end{enumerate}

\section{Related Work}
\label{s:related-work}
\paragraph{Instrument PID Schema.} The PID schema\footnote{\url{https://github.com/rdawg-pidinst/schema/blob/master/schema.rst}} proposed by RDA PIDINST WG~\cite{stocker20pidinst} revolves around providing persistent and unique identifiers to instruments. This ensures that every instrument can be uniquely referenced, thereby aiding in data traceability. The schema encompasses attributes such as the instrument's name, location, type, details of the institution responsible for the instrument. By adhering to a unified schema, it becomes substantially easier to interlink datasets, papers, and instruments, thereby promoting data interoperability.

\paragraph{Scholarly Concepts Extraction.}
Brack et al.\cite{ner_brack} presented the active learning based method which aims at extracting domain-independent scientific concepts from scholarly articles. The approach was applied on 10 different domains including, computer science, engineering, and medicine to show the applicability of the proposed method. Authors presented another approach known as unified deep learning architecture~\cite{sentence_classification} for the classification of sentences obtained from scholarly articles. The approach was tested on the datasets that contain abstracts as well as full articles. D'Souza et al.~\cite{cs_ner} proposed the named entity recognition model that extracts the important entities (research problem, methods, dataset etc.) from computer science articles. Lin et al.~\cite{satellite_ner} applied pre-trained language model (RoBERTA) and distant supervision to the earth observation domain to automatically extract satellite and instrument entities from unstructured texts.

\paragraph{Scientific Knowledge Graphs.}
Knowledge graph technology has been widely adopted to represent the important information of scientific documents in structured form across different domains including healthcare, engineering, education and earth sciences. Zhu el al.~\cite{zhu2021environmental} proposed an approach to represent the environmental observations data in a KG by utilizing LOD principles. Phuoc et al.~\cite{LEPHUOC201625} provided integrated, and live view for heterogeneous IoT data sources using Linked Data, called the Graph of Things (GoT). Wu et al.~\cite{9883498} addressed prevalent challenges in contemporary meteorological research that relies on sensor data for various analytical models. Instead of using the conventional method for combining and preprocessing large blocks of fixed sensor data to generate an enhanced dataset, they accessed remote knowledge graphs to retrieve and augment the sensor data in a scalable manner.

We propose an approach to extract (meta)data of instruments from diverse data sources, and publish it in a knowledge graph. To the best of our knowledge, this is a first attempt to describe instruments information in structured form while maintaining the links to other scholarly artefacts, specifically datasets and articles. The information about instruments and the data produced by these instruments has remained scattered across multiple platforms. This fragmentation poses significant challenges for researchers aiming to have a holistic understanding of instrument functionalities and the data produced, which we aim to address by integrating different artefacts from different sources in a knowledge graph.

\section{Methodology}
\label{s:methodology}
\begin{figure*}[t!]
  \centering
  \includegraphics[width=\textwidth]{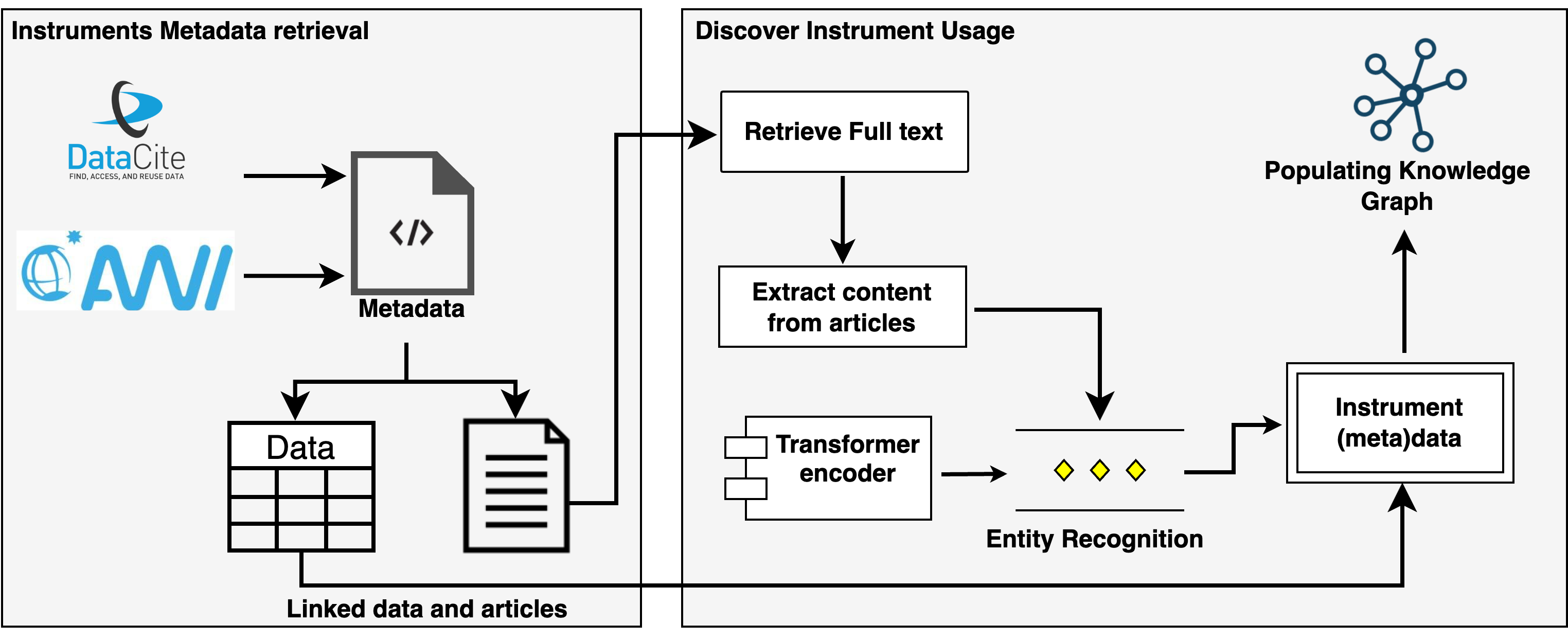}. 
  \caption{Pipeline for populating a knowledge graph using scientific knowledge extracted about instruments: 1) Mining instruments information from DataCite and AWI using GraphQL endpoint and REST API, respectively; 2) Extracting instruments metadata by analyzing the response of APIs; 3) Retrieving artefacts (datasets and articles) linked to instruments; 4) Finding articles that have cited instruments papers and determining the purpose of using instruments using machine learning-based approach; 5) Knowledge graph population with scientific knowledge extracted about instruments.}
  \label{fig1}
\end{figure*}

In this section, we present our methodology for automatically extracting \emph{scientific} knowledge related to instruments and populating a knowledge graph with the extracted (meta)data. Figure~\ref{fig1} provides an overview of the key components: (i) Mining instruments metadata (ii) Retrieve instrument-related artefacts (iii) Discover instruments usage (iv) Populate a knowledge graph.

\subsection{Mining Instruments Information}
There exists several data sources that provide information about instruments, in this paper we consider two, namely (i) DataCite (ii) Alfred Wegener Institut (AWI). DataCite provides extensive, standardized metadata on instruments and research datasets, while the AWI supplies specialized information on instruments used in polar and marine research. The rich and heterogeneous data retrieved from both sources present significant integration challenges, and creates the need for approaches to transform them into a knowledge graph. Note that our method could also be utilized to retrieve instruments information from other sources.

We mine instruments information from DataCite and AWI by leveraging the GraphQL\footnote{\url{https://api.datacite.org/graphql}} and REST API\footnote{\url{https://dashboard.awi.de/data-xxl/api/\#/default/saveData}}, respectively. Then we analyze APIs results to extract metadata about (measuring) instruments, specifically, title, description, manufacturer, and organization. The DataCite query also provide information about scholarly articles that describe the instruments. Since our aim is to incorporate instruments metadata into a knowledge graph, we harmonize the extracted information according to the standard schema defined by the RDA (Research Data Alliance) PIDINST Working Group, being used to describe the instruments information in a standardized way~\cite{stocker20pidinst}.

\subsection{Discover Instruments Usage}
This step entails two tasks. (i) Retrieve articles that mention the use of instruments in research work, as well as datasets produced by these instruments. (ii) Applying transformers-based Named Entity Recognition (NER) approach to extract the important entities from articles to analyze the context of using instruments in research.
\subsubsection{Retrieving Instruments-related Artefacts}
\paragraph{Step 1. Extracting artefacts linked with instruments metadata.}
Some datasets produced by instruments are published and can be globally discovered, while others are only described within articles that discuss the use of these instruments. The latter can be accessed by employing Natural Language Processing (NLP) methods. Published datasets can be retrieved using identifiers (DOIs) or name of instruments. We have retrieved instruments metadata from the DataCite and AWI. To access the data produced by these instruments, we utilize the APIs of DataCite and PANGAEA\footnote{\url{https://www.pangaea.de/}}. Datasets produced by AWI instruments are published on PANGAEA. In contrast, datasets associated with instruments from DataCite are generally retrievable through DataCite itself. By interpreting the APIs response, we collect information about the links between instruments and datasets as well as datasets and articles. Through this step, we infer which datasets are produced by specific instruments.

\paragraph{Step 2. Retrieve articles describing datasets.}
Several articles discuss the data produced by instruments, but such data is not formally assigned a DOI or linked to the corresponding instrument's metadata. Our aim is to retrieve such data by analyzing the content of these articles. We retrieve relevant articles by exploring the citations of papers describing instruments. To locate such papers, first we identify papers that describe instruments (retrieved in step 1 by analyzing the metadata), then we retrieve their citations for further analysis. We analyze the full text of these papers to extract dataset entities and other related scientific entities. This step ensures that information about datasets, which is not formally documented in the metadata, can still be retrieved through the analysis of the relevant content that describe them.

Analyzing articles from both steps 2 and 3, ensures comprehensive information extraction. It effectively captures both explicit and implicit references to datasets, along with additional scientific information. The extracted articles may contain additional contextual information. Analyzing the full text of these papers is crucial to extract further scientific entities such as methods or processes, which are not always explicitly available in the metadata.
\subsection{Analyzing Instruments Usage}
Having retrieved the information about the instruments and associated digital artefacts (including articles and datasets), we now focus on analyzing the content of these artefacts. This analysis aims to extract key parameters measured by instruments. We illustrate this process using a running example involving the CTD instrument, published dataset~\cite{hebbeln2014pofc} and the article by Hebbeln et al.~\cite{bg-11-1799-2014}.
\subsubsection{Analyzing Datasets Content}
In this step, we analyze the content of datasets obtained in Step 1. Our focus is to obtain the parameters measured by these instruments. Specifically, our approach involves a thorough analysis of the datasets content, from which we extract the date and location of experiments as well as key measured parameters. This involves a comprehensive analysis of the datasets to extract the key information. The primary aim here is to uncover the parameters measured by the research instruments. From our example dataset, we extract temporal bounds (start and end dates) (start: 2012-03-21, end: 2012-03-24) and location (Yucatan Strait) of an experiment, as well as key environmental parameters such as salinity, density, and water temperature.

\subsubsection{Analyzing Articles Full text using NER Model}
Although, we have analyzed the datasets content to extract parameters measured by instruments, we extend our approach to include the analysis of articles content (retrieved in step 1 and step 2) that mention the use of instruments. This expanded analysis aims to extract additional details, including datasets (if available) and methods employed. We employ transformer-based language models specifically, BERT, BERT-CRF and BERT-BiLSTM-CRF to discover the context of using instruments in the research work.
\paragraph{\textbf{Data Preparation}}
First, we utilize the Wikiann dataset~\cite{pan2017cross}, which is renowned for its coverage of location entities. Second, we use the STEM-ECR (Scientific Entity Annotations) corpus, which is particularly rich in scholarly entities related to Process, Material, Method, and Data. These entities are defined as follows:
\begin{itemize}
    \item Method: Techniques, tools, or procedures used for data collection or analysis.
    \item Data: Specific observations, measurements, or numerical values reported in the study.
    \item Process: Practices employed to uncover knowledge and interpret the meaning of those discoveries.
    \item Material: Physical materials or substances studied or used within the experimental setup.
\end{itemize}

The integration of two datasets ensures that our model is well-exposed to a variety of entity types that are important to our work. Given the limited number of entries in STEM-ECR corpus--- we leverage Large Language Model (LLM), specifically GPT-3 to generate synthetic data. This is achieved by feeding GPT-3 the existing data along with crafted prompts that guide the model to generate additional entries for Data, Method, and Process entities, thereby increasing their representation in the context required for our NER tasks. This richer dataset substantially improves the accuracy of our NER models and facilitate more effective identification and classification of relevant entities across varied content.
\paragraph{\textbf{Model Training}}
We evaluate the performance of three advanced transformer-based models for the task of Named Entity Recognition (NER): BERT, BERT-CRF, BERT-BiLSTM-CRF. Each model was trained and tested on the above mentioned dataset. BERT (Bidirectional Encoder Representations from Transformers), known for its effectiveness in capturing context within language; BERT-CRF (Conditional Random Fields), which combines BERT's deep learning capabilities with CRF's sequence modeling to enhance entity recognition. We compare the performance of these models and use results of the model that provides high accuracy. Our goal is to identify these entities to discover the context of using instruments. The computations were carried out on Mac M2 GPU.

BERT and BERT-CRF models are useful to generate powerful token-level representations that captures the important meanings and relationships within the text. During training, we employed the Adam optimizer~\cite{kingma2014adam} and our training and validation data were batched with a batch size of 16 and the model was trained for 25 epochs. We then integrate CRF layer with BERT model for sequence modeling tasks. CRF is particularly used for predicting the tags or labels of tokens (method, dataset, and location) in a sequence. The integration of CRF with BERT brings together the deep contextualized embeddings from BERT and the sequential modeling advantages of CRF. This fine-tuning process allows us to leverage the capabilities of the pretrained BERT model, while adapting it to recognize the specific entities relevant to our research.

Through comparative analysis, we identify model that yields the best results for different entity types (Data, Material, Method, Process and Location) entities. To assess each model's effectiveness, we used standard NER metrics including precision, recall, and the F1-score. In our case, BERT-CRF shows better results (Table \ref{table1}).

\paragraph{\textbf{Retrieve Scientific Entities from Articles Text}}
To determine the context of using instruments, we systematically analyze the full text of articles retrieved using the Unpaywall REST API\footnote{\url{https://api.unpaywall.org/v2/10.1186/s12920-019-0613-5?email=unpaywall\_01@example.com}}. Unpaywall provides the URL to the PDF version of scholarly articles. We employ the trained NER model to extract specific entities, specifically, datasets, methods, processes and location from the full text of articles. From our running example, we retrieve the backscatter as data, and hydroacoustic measurements, and water column studies, as processes.

\begin{lstfloat}
\begin{lstlisting}[language=XML, caption={Triple statements < \textit{subject, predicate, object} > that are ingested into the ORKG to enrich the knowledge graph. The listing includes information about the instrument, and produced datasets, and shows the relationships among these entities}., basicstyle=\ttfamily\scriptsize, numbersep=5pt, showspaces=false, xleftmargin=0.5cm, showstringspaces=true, label=listing1, frame=tb]
# paper mentioning instrument (CTD)
paper=Environmental forcing of the Campeche cold-water coral
province, southern Gulf of Mexico
dataset = Physical oceanography from CTS during maria S. Merain
instrument_devices = CTD RBR, CTD_Seabird-SBE-19-0
parameters = depth water, salinity

# triples describing paper contributions in ORKG
< paper, (*@\textit{contribution}@*),  experimentDetails >
< experimentDetails, (*@\textit{data}@*), dataset >

< dataset, (*@\textit{producedBy}@*), CTD>
< experimentDetails, (*@\textit{parameters}@*), parameters >
< CTD, (*@\textit{devices}@*), instrument_devices >

\end{lstlisting}
\end{lstfloat}

\subsection{Knowledge Graph Population}
In this step, we leverage the Open Research Knowledge Graph (ORKG) to ingest the retrieved (meta)data about instruments. We form triples using instrument-related data and add them into the ORKG.
\subsubsection{Identify Research Fields of Articles}
To add articles in the ORKG, it is essential to link them with their respective research fields\footnote{\url{https://orkg.org/help-center/article/20/ORKG\_Research\_fields\_taxonomy}}, for example, information systems, database systems, etc. To identify the research fields of articles, we utilize a zero-shot classification technique to automatically categorize research articles based on their titles and abstracts into predefined research fields. The list of research fields was sourced from the ORKG, to ensure that our classification aligns with established categories/research fields. We opt for zero-shot learning, a method that does not require labeled training data. We employ the GPT-3 LLM to classify the research fields of articles in zero-shot setting. By feeding the model with the articles titles and abstracts along with a set of research fields labels from the ORKG, the model assesses the relevance of each text according to the provided labels. It then outputs the most relevant research field.
\subsubsection{Ingest Instruments (Meta)data into the KG.}
We now enrich the knowledge graph with the scientific knowledge obtained in the analysis of instruments (meta)data. For this, we leverage the Open Research Knowledge Graph (ORKG). We follow the ORKG ontology\footnote{\url{https://gitlab.com/TIBHannover/orkg/orkg-ontology/-/blob/master/orkg-core.ttl?ref\_type=heads}} to describe the extracted data. The ORKG aims to represent the essential information published in scholarly articles in a machine actionable and structured form. The ORKG represents research contributions describing key results, the materials and methods used to obtain the results, and the addressed research problem. In order to integrate the heterogeneous instruments (meta)data into a knowledge graph, it is necessary to express it into a consistent format. The RDA PIDINST Working Group provides a standardized schema, that allows to describe instruments metadata in a harmonized way. We utilize that schema to standardize the metadata of instruments retrieved from different sources, and using that schema we add the metadata of instruments in ORKG.

\begin{figure*}[t!]
  \includegraphics[width=\textwidth]{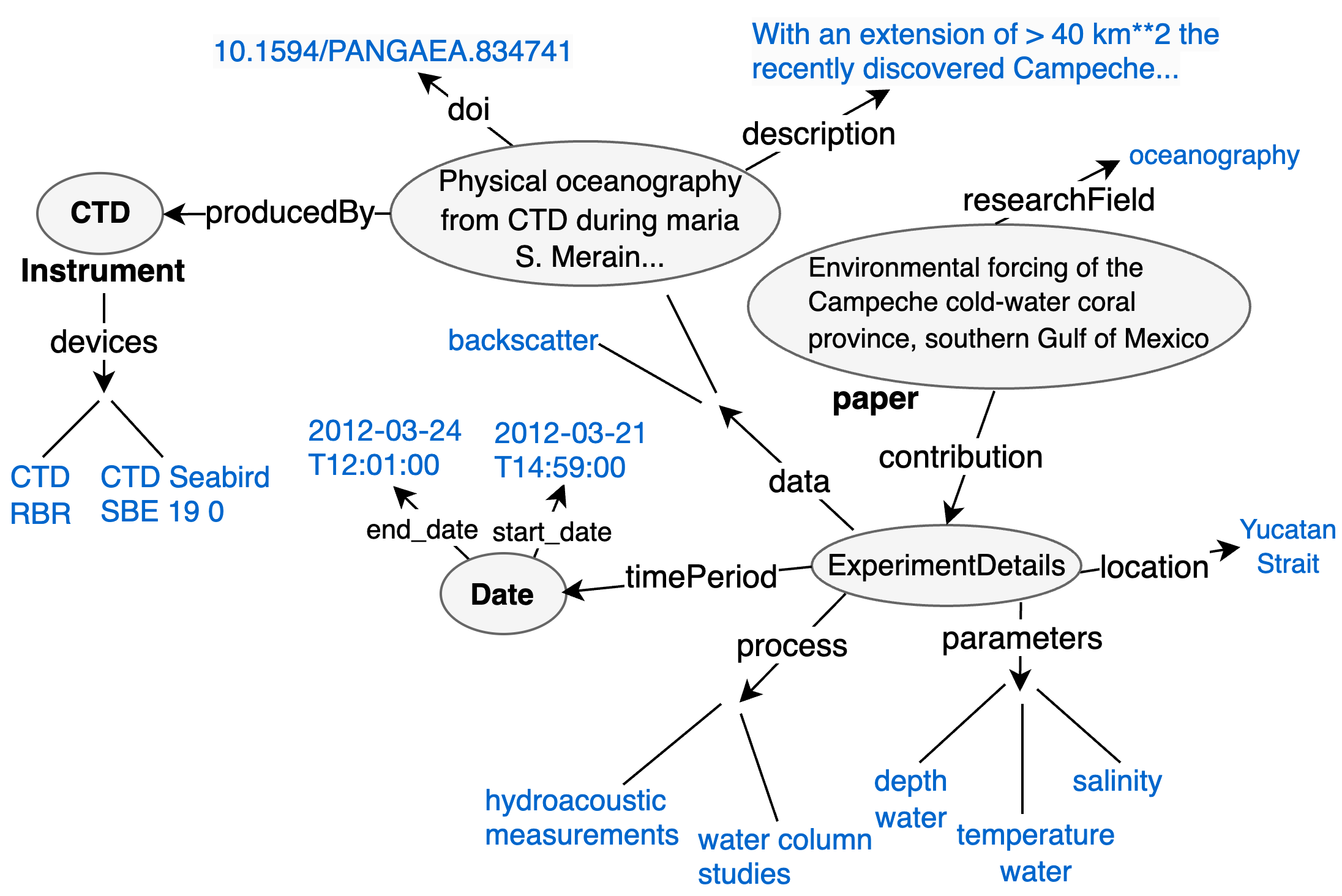}
  \caption{The subgraph of the instrument (CTD), which has been constructed by extracting the data from multiple data sources; Showing the information of the instrument, produced data, and other details (processes and measured parameters) extracted from the dataset content and the full text of the article using the NER model. (In ORKG, each paper is described in the form of research contributions, as show in Listing\ref{listing1}).}
  \label{fig4}
\end{figure*}
The scientific information extracted from the analysis of (meta)data of instruments is organized in triples (Listing~\ref{listing1}) and ingested into ORKG using its REST API. The API is called iteratively to add the papers, datasets and the instrument metadata into the ORKG. Figure~\ref{fig4} shows the graph created for an instrument (CTD) by extracting the linked datasets and articles from different data sources. It shows the diverse information about an instrument, including instrument details, datasets produced by the instrument, as well as the article that mention the use of this instrument. While enriching the knowledge graph, the extracted scientific knowledge (datasets, methods, and processes) is represented as research contributions in the ORKG.

\subsection{Knowledge Graph Usage}
This section showcases the practical application of the enriched knowledge graph. By means of visualization and querying, we underscore the potential of our knowledge graph as a pivotal tool for researchers aiming to gain detailed information about instruments and their associated datasets.

\paragraph{Knowledge visualization.}
The knowledge graph can visually show the connections between instruments, datasets, and articles. Details of instruments and other related artefacts, which are scattered across different data sources, can be interactively explored through the front-end module of ORKG. This will potentially help users to understand the context in which instruments are used in research. Users can select an instrument to further explore information, such as which articles mention its usage and which datasets have been produced by the instrument.

\paragraph{Querying Knowledge Graph.}
We execute different queries on the knowledge graph to retrieve interconnected details about instruments. Listing~\ref{listing2} shows different SPARQL queries executed to retrieve instrument-related information.

\lstset{escapeinside={(*@}{@*)}}
\begin{lstfloat}[t!]
\begin{lstlisting}[language=XML, caption=SPARQL queries executed on RDF data source to retrieve instrument-related artefacts., basicstyle=\ttfamily\scriptsize, mathescape=true, numbersep=5pt, showspaces=false, xleftmargin=0.5cm, showstringspaces=true, label=listing2, frame=tb]
(*@\textbf{Query 1}@*): SPARQL query used to obtain the artefacts related to E2 -
Flat-Cone Diffractometer instrument
(*@\rule{\linewidth}{1pt}@*)
PREFIX orkgr: <http://orkg.org/orkg/resource/>
PREFIX orkgc: <http://orkg.org/orkg/class/>
PREFIX orkgp: <http://orkg.org/orkg/predicate/>
PREFIX rdfs: <http://www.w3.org/2000/01/rdf-schema#>
PREFIX xsd: <http://www.w3.org/2001/XMLSchema#>
PREFIX rdf: <http://www.w3.org/1999/02/22-rdf-syntax-ns#>

SELECT ?paper_title ?dataset ?instrument_name
    WHERE {
        ?paper rdf:type orkgc:Paper;
           rdfs:label ?paper_title;
           orkgp:P31 ?contribution.
    ?contribution orkgp:P4017 ?object.
    ?object orkgp:P146018 ?instrument.
    ?object rdfs:label ?dataset.
    ?instrument rdfs:label ?instrument_name.
    FILTER(?instrument = <http://orkg.org/orkg/resource/R741211>) }
(*@\rule{\linewidth}{1pt}@*)
(*@\textbf{Query 2}@*): Retrieving all papers and associated data collected using Box corer instrument in the Arctic Ocean (ORKG resource ID: R694251)
(*@\rule{\linewidth}{1pt}@*)

SELECT ?paper_title ?dataset_label ?instrument_name ?sea
    WHERE {
        ?paper rdf:type orkgc:Paper;
           rdfs:label ?paper_title;
           orkgp:P31 ?contribution.

    ?contribution orkgp:P2005 ?dataset.
    ?contribution orkgp:P5049 ?location.
    ?dataset rdfs:label ?dataset_label;
             orkgp:P146018 ?instrument.
      
    ?instrument rdfs:label ?instrument_name.
    ?location rdfs:label ?sea.
    # URI to Box corer
    FILTER(?instrument = <http://orkg.org/orkg/resource/R694631> &&
        # URI to Arctic Ocean
        ?location = <http://orkg.org/orkg/resource/R694251>) }

(*@\rule{\linewidth}{1pt}@*)
(*@\textbf{Query 3}@*): Which instruments are most commonly used to measure temperature and salinity parameters in Arctic ocean?
(*@\rule{\linewidth}{1pt}@*)
    SELECT ?instrument ?parameters ?location_name
    WHERE {
        ?instrument rdf:type orkgc:C99025;
                    rdfs:label ?label.
    ?subject orkgp:P146018 ?instrument.
    ?contribution ?predicate ?subject;
                  orkgp:P5049 ?location.
    ?location rdfs:label ?location_name.
    Optional { ?contribution orkgp:P15680 ?parameters. }
    FILTER ((CONTAINS(?parameters, "Temperature") || CONTAINS(?parameters, "Salinity")) &&
        ?location = <http://orkg.org/orkg/resource/R694251>
    )}
\end{lstlisting}
\end{lstfloat}

The above-mentioned queries retrieve all datasets produced by the specified instruments, as well as all articles that cite this instrument. The results also include the article descriptions (dataset, method and location entities) which are retrieved by analyzing the content of articles (retrieved in step 2) using the NER model.

\section{Results Evaluation}
\paragraph{Performance evaluation of models.} We evaluated our NER models on the test datasets from STEM-ECR and Wikiann. The results are shown in Table~\ref{table1}. We evaluated the model's performance using standard metrics such as Precision, Recall, and F1-Score. These scores indicate that the BERT-CRF model after fine-tuning produced reliable results.

\begin{table}[t!]
\centering
\caption{Performance Evaluation of NER Model}
\label{table1}
\begin{tabular}{l|c|c|c|c|c|c|c|c|c|c|c|c|c|c}
\hline
\textbf{Model} & \multicolumn{3}{c|}{\textbf{BERT}} & \multicolumn{3}{c|}{\textbf{BERT-CRF}} & \multicolumn{3}{c|}{\textbf{BERT-BiLSTM-CRF}} \\
\hline
Class & F1 & Precision & Recall & F1 & Precision & Recall & F1 & Precision & Recall \\
\hline
\textbf{Location} & 0.83 & 0.82 & 0.83 & 0.92 & 0.89 & 0.94 & 0.90 & 0.89 & 0.93  \\
\textbf{Data} & 0.64 & 0.62 & 0.65 & 0.77 & 0.75 & 0.78 & 0.77 & 0.73 & 0.82  \\
\textbf{Method} & 0.81 & 0.75 & 0.89 & 0.85 & 0.82 & 0.88 & 0.86 & 0.85 & 0.88  \\
\textbf{Process}  & 0.61 & 0.60 & 0.62 & 0.69 & 0.70 & 0.68 & 0.68 & 0.68 & 0.68 \\
\textbf{Material} & 0.68 & 0.68 & 0.68 & 0.83 & 0.83 & 0.83 & 0.83 & 0.83 & 0.83 \\
\hline
\end{tabular}
\end{table}

\paragraph{KG Data Quality.}
The information about instruments and datasets is dispersed across different data sources, therefore we verify whether the relations among these entities are correct and logical. The links among articles, datasets and instruments have been completely checked by crowd sourcing, and their correctness can be ensured. Specifically, links among instruments to datasets are thoroughly reviewed because these links must be established correctly to ensure that the fragmented information is linked correctly and can be retrieved easily.

Our main goal was to discover the use of instruments in the research work. It is essential to determine that the proposed approach reliably produces the desired results. We verify the correctness of the results extracted using the NER model. Although we have achieved better accuracy on the test dataset, but manual verification of the ORKG paper descriptions is also important. This manual verification involves examining a sample of articles that cite instruments, then the descriptions of these articles are thoroughly reviewed by comparing them with the information published in original articles. Our analysis shows that the data added in the ORKG is of high quality. Since the ORKG supports crowd sourcing, the machine-readable descriptions of papers can be further enriched with the help of domain experts.

\begin{table}[t!]
\centering
\caption{Statistics about the artefacts derived from the analysis of instruments (meta)data also showing the total statements added into the ORKG.}
\label{table2}
\begin{tabular}{l|l}
Entity & Count \\ \hline
Instruments & 131 \\ \hline
Instruments from Datacite & 46 \\ \hline
Instruments from AWI & 85 \\ \hline
Datasets produced by Instruments & 51,952 \\ \hline
Articles linked with datasets & 4,345 \\ \hline
\end{tabular}
\end{table}

\section{Discussion}
\label{s:discussion}
Table~\ref{table2} presents a comprehensive overview of various research entities along with their corresponding quantities. Specifically, there are a total of 131 instruments, with 46 mined from Datacite and 85 from AWI. A total of 51,952 datasets have been produced by these instruments and the information of these datasets is taken from DataCite and PANGAEA. Moreover, these instruments have been used in research work to produce the data which is described in research articles. Information about these datasets is fetched from various articles using the NER model. Additionally, 4,345 articles are linked to datasets.


The ORKG is a domain-agnostic scholarly infrastructure, and enables machine-readable descriptions of scholarly knowledge across a wide range of research domains. We have leveraged this functionality and further enriched it with articles published about Oceanography and Materials Science and Engineering research fields. Thus, with our approach, we have automatically curated two research fields in the ORKG.

Our work is an important step towards integrating instruments-related information into the ORKG. The enriched KG contains instruments metadata, data produced by them as well as articles which mention the use of these instruments in research work. As before, such information is scattered across different data sources, its navigation is cumbersome and some useful information can remain undiscovered. With our approach, the useful links can be uncovered which are not straightforward in its current form. Consider the scenario where a specific instrument is referenced across numerous articles. Extracting detailed information about its usage becomes a tedious task. Initially, one must identify the articles mentioning its usage in a research work which is achievable either by searching databases with the instrument name or by exploring the citations of the paper that describes the instrument. Following this, each identified article must be manually checked to understand the context in which the instrument was used. This multi-step process is a cumbersome activity for gather comprehensive usage information about a particular instrument. Retrieving data from various sources, produced by instruments, and integrating it into a knowledge graph is a crucial step for the quick discovery of the contextual information associated with these instruments.

\paragraph{Future Directions.} We aim to retrieve information about instruments from other data sources which will potentially increase the instruments, datasets and articles information in the ORKG. We also aim to develop a question answering system, where users can pose their queries in natural language text to retrieve the details about instruments.

\section{Conclusions}
\label{s:conclusion}
We have presented an approach that retrieves instruments metadata and related digital artefacts (articles and datasets) from multiple data sources, and integrates them into the Open Research Knowledge Graph (ORKG). Our methodology extends beyond mere metadata organization and represents the essential information of articles and datasets in a structured form. Furthermore, linking instruments, with data and articles, provide a comprehensive overview of their roles in driving discoveries across numerous fields. Such an information in a knowledge graph would allow researchers to navigate the instrument-related information more effectively. We believe that our work is an important step towards best practices of research data management as it transforms the distributed information into a unified knowledge base which ultimately enhances the accessibility of scientific information.

\section*{Acknowledgment}
This work was co-funded by the European Research Council for the project ScienceGRAPH (Grant agreement ID: 819536) and the German Research Foundation (DFG) project NFDI4DS (PN: 460234259).

\bibliographystyle{splncs04}
\bibliography{paper}

\begin{thebibliography}{10}
\providecommand{\url}[1]{\texttt{#1}}
\providecommand{\urlprefix}{URL }
\providecommand{\doi}[1]{https://doi.org/#1}

\bibitem{balch1984vertical}
Balch, A.H., Lee, M.W.: Vertical seismic profiling: technique, applications,
  and case histories  (1984)

\bibitem{ner_brack}
Brack, A., D'Souza, J., Hoppe, A., Auer, S., Ewerth, R.: Domain-independent
  extraction of scientific concepts from research articles. In: Jose, J.M.,
  Yilmaz, E., Magalh{\~a}es, J., Castells, P., Ferro, N., Silva, M.J., Martins,
  F. (eds.) Advances in Information Retrieval. pp. 251--266. Springer
  International Publishing, Cham (2020)

\bibitem{sentence_classification}
Brack, A., Hoppe, A., Buscherm\"{o}hle, P., Ewerth, R.: Cross-domain multi-task
  learning for sequential sentence classification in research papers. In:
  Proceedings of the 22nd ACM/IEEE Joint Conference on Digital Libraries. JCDL
  '22, Association for Computing Machinery, New York, NY, USA (2022).
  \doi{10.1145/3529372.3530922}

\bibitem{ctd}
Brown, N.: New generation ctd system (conductivity-temperature-depth sensor)
  (1988). \doi{10.1109/48.567}

\bibitem{cs_ner}
D'Souza, J., Auer, S.: Computer science named entity recognition in the open
  research knowledge graph. In: Tseng, Y.H., Katsurai, M., Nguyen, H.N. (eds.)
  From Born-Physical to Born-Virtual: Augmenting Intelligence in Digital
  Libraries. pp. 35--45. Springer International Publishing, Cham (2022)

\bibitem{enders2020conceptual}
Enders, M., Havemann, F., Ruland, F., Bernard-Verdier, M., Catford, J.A.,
  G{\'o}mez-Aparicio, L., Haider, S., Heger, T., Kueffer, C., K{\"u}hn, I.,
  et~al.: A conceptual map of invasion biology: Integrating hypotheses into a
  consensus network. Global Ecology and Biogeography  \textbf{29}(6),  978--991
  (2020)

\bibitem{knowlife}
Ernst, P., Meng, C., Siu, A., Weikum, G.: Knowlife: A knowledge graph for
  health and life sciences. pp. 1254--1257 (03 2014).
  \doi{10.1109/ICDE.2014.6816754}

\bibitem{bg-11-1799-2014}
Hebbeln, D., Wienberg, C., Wintersteller, P., Freiwald, A., Becker, M., Beuck,
  L., Dullo, C., Eberli, G.P., Glogowski, S., Matos, L., Forster, N.,
  Reyes-Bonilla, H., Taviani, M.: Environmental forcing of the campeche
  cold-water coral province, southern gulf of mexico. Biogeosciences
  \textbf{11}(7),  1799--1815 (2014). \doi{10.5194/bg-11-1799-2014},
  \url{https://bg.copernicus.org/articles/11/1799/2014/}

\bibitem{hebbeln2014pofc}
{Hebbeln}, D., {Wienberg}, C., {Wintersteller}, P., {Freiwald}, A., {Becker},
  M., {Beuck}, L., {Dullo}, W.C., {Eberli}, G.P., {Glogowski}, S., {Matos}, L.,
  {Foster}, N., {Reyes-Bonilla}, H., {Taviani}, M., {Expedition MSM20/4
  Scientists}: {Physical oceanography from CTD during Maria S. Merian cruise
  MSM20/4 in spring 2012} (2014). \doi{10.1594/PANGAEA.834741}, supplement to:
  Hebbeln, D et al. (2014): Environmental forcing of the Campeche cold-water
  coral province, southern Gulf of Mexico. Biogeosciences, 11(7), 1799-1815,
  https://doi.org/10.5194/bg-11-1799-2014

\bibitem{heger2013conceptual}
Heger, T., Pahl, A.T., Botta-Duk{\'a}t, Z., Gherardi, F., Hoppe, C., Hoste, I.,
  Jax, K., Lindstr{\"o}m, L., Boets, P., Haider, S., et~al.: Conceptual
  frameworks and methods for advancing invasion ecology. Ambio  \textbf{42}(5),
   527--540 (2013)

\bibitem{domainspecific}
Jain, N.: Domain-specific knowledge graph construction for semantic analysis.
  In: Harth, A., Presutti, V., Troncy, R., Acosta, M., Polleres, A.,
  Fern{\'a}ndez, J.D., Xavier~Parreira, J., Hartig, O., Hose, K., Cochez, M.
  (eds.) The Semantic Web: ESWC 2020 Satellite Events. pp. 250--260. Springer
  International Publishing, Cham (2020)

\bibitem{kingma2014adam}
Kingma, D.P., Ba, J.: Adam: A method for stochastic optimization. arXiv
  preprint arXiv:1412.6980  (2014)

\bibitem{LEPHUOC201625}
Le-Phuoc, D., {Nguyen Mau Quoc}, H., {Ngo Quoc}, H., {Tran Nhat}, T.,
  Hauswirth, M.: The graph of things: A step towards the live knowledge graph
  of connected things. Journal of Web Semantics  \textbf{37-38},  25--35
  (2016). \doi{https://doi.org/10.1016/j.websem.2016.02.003}

\bibitem{dbpedia}
Lehmann, J., Isele, R., Jakob, M., Jentzsch, A., Kontokostas, D., Mendes, P.,
  Hellmann, S., Morsey, M., Van~Kleef, P., Auer, S., Bizer, C.: Dbpedia - a
  large-scale, multilingual knowledge base extracted from wikipedia. Semantic
  Web Journal  \textbf{6} (01 2014). \doi{10.3233/SW-140134}

\bibitem{satellite_ner}
Lin, M., Jin, M., Liu, Y., Bai, Y.: Satellite and instrument entity recognition
  using a pre-trained language model with distant supervision. International
  Journal of Digital Earth  \textbf{15},  1290--1304 (12 2022).
  \doi{10.1080/17538947.2022.2107098}

\bibitem{chlorophyll_fluorescence}
Maxwell, K., Johnson, G.N.: {Chlorophyll fluorescence—a practical guide}.
  Journal of Experimental Botany  \textbf{51}(345),  659--668 (04 2000).
  \doi{10.1093/jexbot/51.345.659}

\bibitem{pan2017cross}
Pan, X., Zhang, B., May, J., Nothman, J., Knight, K., Ji, H.: Cross-lingual
  name tagging and linking for 282 languages. In: Proceedings of the 55th
  Annual Meeting of the Association for Computational Linguistics (Volume 1:
  Long Papers). pp. 1946--1958 (2017)

\bibitem{stocker20pidinst}
Stocker, M., Darroch, L., Krahl, R., Habermann, T., Devaraju, A., Schwardmann,
  U., D’Onofrio, C., Häggström, I.: Persistent identification of
  instruments. Data Science Journal  \textbf{19} (2020).
  \doi{10.5334/dsj-2020-018}, \url{http://dx.doi.org/10.5334/dsj-2020-018}

\bibitem{stocker2023fair}
Stocker, M., Oelen, A., Jaradeh, M.Y., Haris, M., Oghli, O.A., Heidari, G.,
  Hussein, H., Lorenz, A.L., Kabenamualu, S., Farfar, K.E., et~al.: Fair
  scientific information with the open research knowledge graph. FAIR Connect
  \textbf{1}(1),  19--21 (2023)

\bibitem{wikidata}
Vrande\v{c}i\'{c}, D., Kr\"{o}tzsch, M.: Wikidata: A free collaborative
  knowledgebase. Commun. ACM  \textbf{57}(10),  78–85 (sep 2014).
  \doi{10.1145/2629489}, \url{https://doi.org/10.1145/2629489}

\bibitem{9883498}
Wu, J., Orlandi, F., Pathan, M.S., O'Sullivan, D., Dev, S.: Augmenting weather
  sensor data with remote knowledge graphs. In: IGARSS 2022 - 2022 IEEE
  International Geoscience and Remote Sensing Symposium. pp. 1264--1267 (2022).
  \doi{10.1109/IGARSS46834.2022.9883498}

\bibitem{zhu2021environmental}
Zhu, R., Stephen, S., Zhou, L., Shimizu, C., Cai, L., Mai, G., Janowicz, K.,
  Hitzler, P., Schildhauer, M.: Environmental observations in knowledge graphs.
  In: DaMaLOS. pp. 1--11 (2021)

\end{thebibliography}
\end{document}